# Reviewing definition of resilience in different disciplines with a focus on disaster restructure systems

Saviz Saei
ss4646@msstate.edu

Nazanin Tajik
tajik@ise.msstate.edu

## Abstract

A key principle in resilience thinking is" Embracing Change" because change is, indeed, inevitable. In the face of a growing number of disasters, natural and human-made disasters, our critical infrastructures (CIs) are being challenged like never before. This recent trend has sparked a wave of interest among both practitioners and researchers in understanding and delving deeper into the concept of resilience across multiple disciplines. This paper provides an accessible review of these new insights, exploring various frameworks, guidebooks, and methodologies that define resilience through the lens of ecology, engineering, psychology, social science, community, and disaster management during crisis.

## 1. Introduction

Resilience is a multifaceted construct that has different meanings across social science, engineering, psychology, and community studies. How resilience is defined and measured across these fields can lead to overlapping and contrasting interpretations. The challenge in defining resilience is prominent due to its varied interpretations across disciplines. Yet, the essence is universally understood as the system's ability to endure, adapt, and recover from disruptions. This systematic review highlights the need for future research to create a unified, multidimensional resilience framework, incorporating perspectives from entities like the National Institute of Standards and Technology (NIST), the Joint Research Centre (JRC)-European Commission's in-house scientific research center, and the Federal Emergency Management Agency (FEMA), an agency of the United States Department of Homeland Security. The collective findings underscore the complexity and multifaceted nature of resilience, leading to the realization that a unified and multidimensional framework is essential. Such a framework would integrate the various interpretations of resilience across different disciplines and provide a more holistic approach to understanding and enhancing the resilience of infrastructure systems. The systematic review conducted in this study sheds light on the current state of resilience research and pinpoints the gaps that persist. It calls attention to the need for a cohesive framework to bridge these gaps, backed by the extensive, multidisciplinary body of literature that spans the community to psychological resilience underpinned by solid engineering principles. The review paves the way for future research to build upon the existing knowledge base, drawing from the rich insights provided by leading institutions and the thorough analysis of the literature to forge pathways that will fortify the resilience of infrastructure systems against the diverse array of challenges they face. Most articles fall under the 'Quantification Model' category, particularly within the Engineering discipline, accounting for 28% of the total. This underscores the technical orientation of the field, where precise modeling and analytical methodologies are paramount. However, the 'Conceptual Framework' category also showcases a robust engagement with the Social Sciences and Disaster Management sectors, capturing 21% and 17%, respectively, which implies a recognition of the broader social and systemic implications of resilience.

## 2. Review Methodology

A systematic review is an approach that minimizes the occurrence of systematic and random errors  (Mahtani et al., 2020). To review the literature "systematically," the collection of published articles regarding the resiliency of infrastructure systems is captured by Google Scholar, Elsevier, IEEE Xplore, and Nature in conjunction with electronic library records. Figure 1 illustrates the flowchart of the review process, which follows the three steps:

- Step 1: Identify resources and keyword searches
- Step 2: Assess the quality of articles
- Step 3: Summarize the findings and analysis

This study also explores existing gaps and challenges in the current research and proposes future directions to enhance resilience in these crucial systems. It also leverages the reports and research findings from esteemed institutions such as NIST, JRC, and FEMA to delve into existing gaps and challenges in the field of resilience. It integrates insights from these organizations to propose future directions for enhancing resilience in crucial systems, encompassing community, ecosystem, social factors, and engineering. Step 1 conducts a keyword search limited by adding one keyword after another. Search keys included a combination of "resilience", "community", "ecosystem", "social", "psychology", and "engineering". The process of searching is not limited to the search keys in the title and abstract but includes all the articles' text. The result led to 2,590,000 engineering, 407,000 ecosystems, 20,000 psychological, and 2,620,000 social and community resilience. Then, the articles excluded the areas of "vulnerability", "recovery", and extracted the areas other than a disaster. However, In the quest for resilience assessment methods, various terms such as "tool", "toolkit", "model", "framework", "guidebook", "review", and “index” have been utilized. Step 2 implements the results based on the definition, review, concept, methodology, and solution approach of resilience as the double screening approach, and results in 120 articles. The consequence of the systematic search implementation led to 28 publications being retrieved in peer-reviewed journals, conference proceedings, and survey research. The result of the collected articles was reported by 110 different journals, mostly in “The International Journal of Disaster Risk Reduction”, “Natural Hazards”, and “Reliability Engineering & System Safety” journals. Table 1 shows the summary of the reviewed articles.

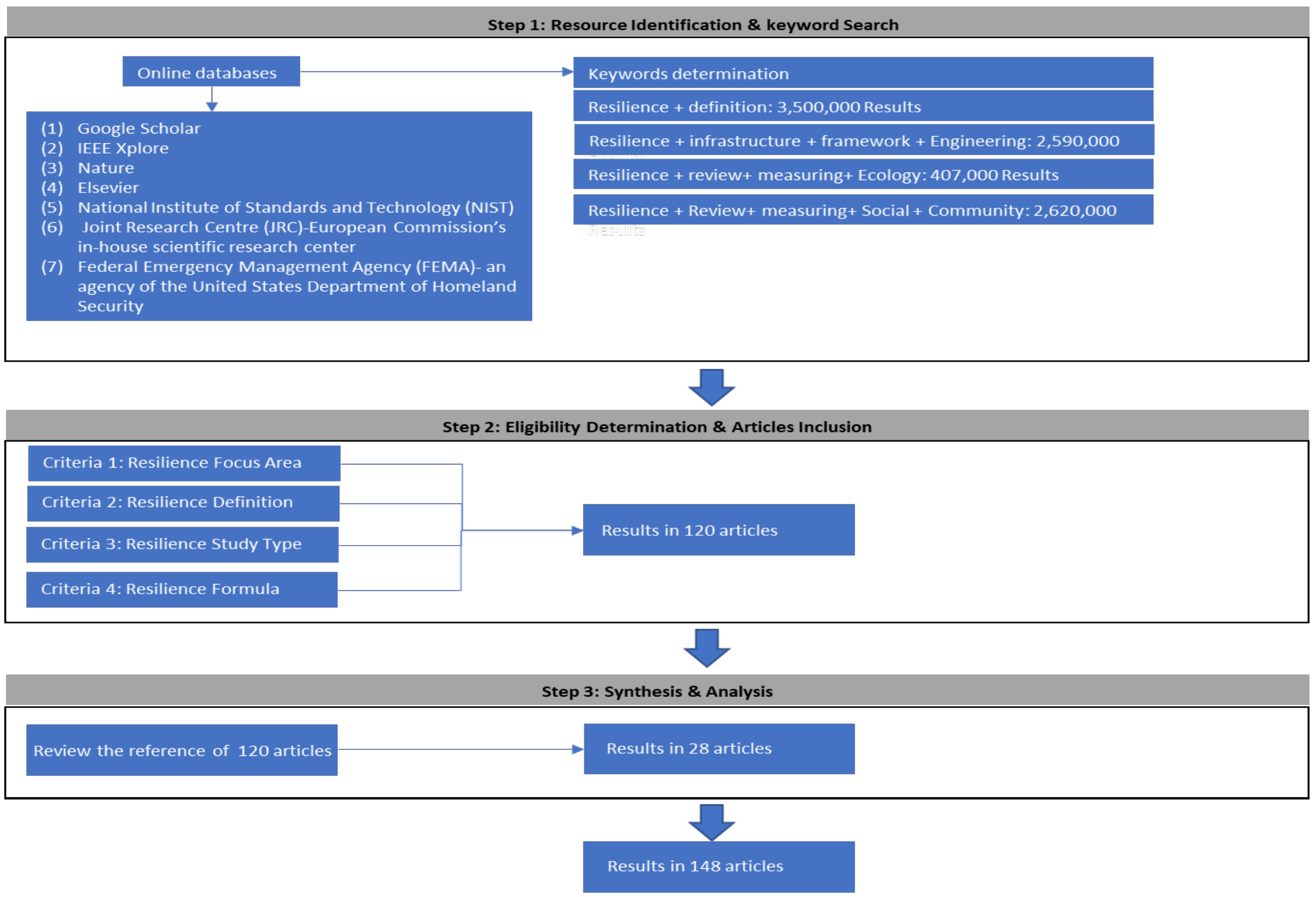

Figure 1. Flowchart of the literature review process on resilience

The challenge in defining resilience is prominent due to its varied interpretations across disciplines. Yet, the essence is universally understood as the system's ability to endure, adapt, and recover from disruptions. This systematic review highlights the need for future research to create a unified, multidimensional resilience framework, incorporating perspectives from entities like NIST, JRC, and FEMA. The collective findings underscore the complexity and multifaceted nature of resilience, leading to the realization that a unified and multidimensional framework is essential. Such a framework would integrate the various interpretations of resilience across different disciplines and provide a more holistic approach to understanding and enhancing the resilience of infrastructure systems. The systematic review conducted in this study sheds light on the current state of resilience research and pinpoints the gaps that persist. It calls attention to the need for a cohesive framework to bridge these gaps, backed by the extensive, multidisciplinary body of literature that spans community to psychological resilience underpinned by solid engineering principles. The review paves the way for future research to build upon the existing knowledge base, drawing from the rich insights provided by leading institutions and the thorough analysis of the literature to forge pathways that will fortify the resilience of infrastructure systems against the diverse array of challenges they face. The systematic review addresses key questions to enhance the understanding of infrastructure resilience. The questions this review aims to answer are as follows:

- What constitutes 'resilience' across different disciplines, and how is it defined in the context of infrastructure systems? This question probes the varied interpretations of resilience and aims to clarify its meaning within the scope of infrastructure.
- What are the existing methods for assessing the resilience of infrastructure systems? The

review examines the tools, models, frameworks, and indices developed to evaluate resilience.

- To what extent do these methods capture the multidimensional nature of resilience? This question scrutinizes the comprehensiveness of current methods in encapsulating the complexity of resilience.

- How do current research and methods address the interplay between engineering, community, ecosystem, and social factors in resilience? The review assesses how well the current body of research integrates these varied aspects into a unified approach to resilience.

## 3. Review Statistics

Figures 2 and 3 display the distribution and interrelationships of scholarly work across disciplines. Figure 2 illustrates the percentage of review, framework, guide books, statistical analysis, concept, mathematical formulation, methodology, and report view the resilience through different disciplines, such as General (GE), Engineering (Eng), Ecology (Eco), Social Science (SSc), Psychology (Psy), Community (CM), and Disaster Management (DM). Engineering predominantly utilizes methodologies and formulas, while Disaster Management shows a balanced integration of different scholarly outputs, reflecting its interdisciplinary nature. Social Science and Psychology appear to emphasize conceptual work, indicative of their qualitative focus, which provides a foundational understanding crucial for further research. Conversely, Ecology is strongly associated with statistical analysis. While some disciplines may lean towards qualitative analysis in establishing foundational knowledge, others employ more quantitative methods, offering concrete tools for measuring resilience, which are critical in areas such as policy development, system engineering, and management practice

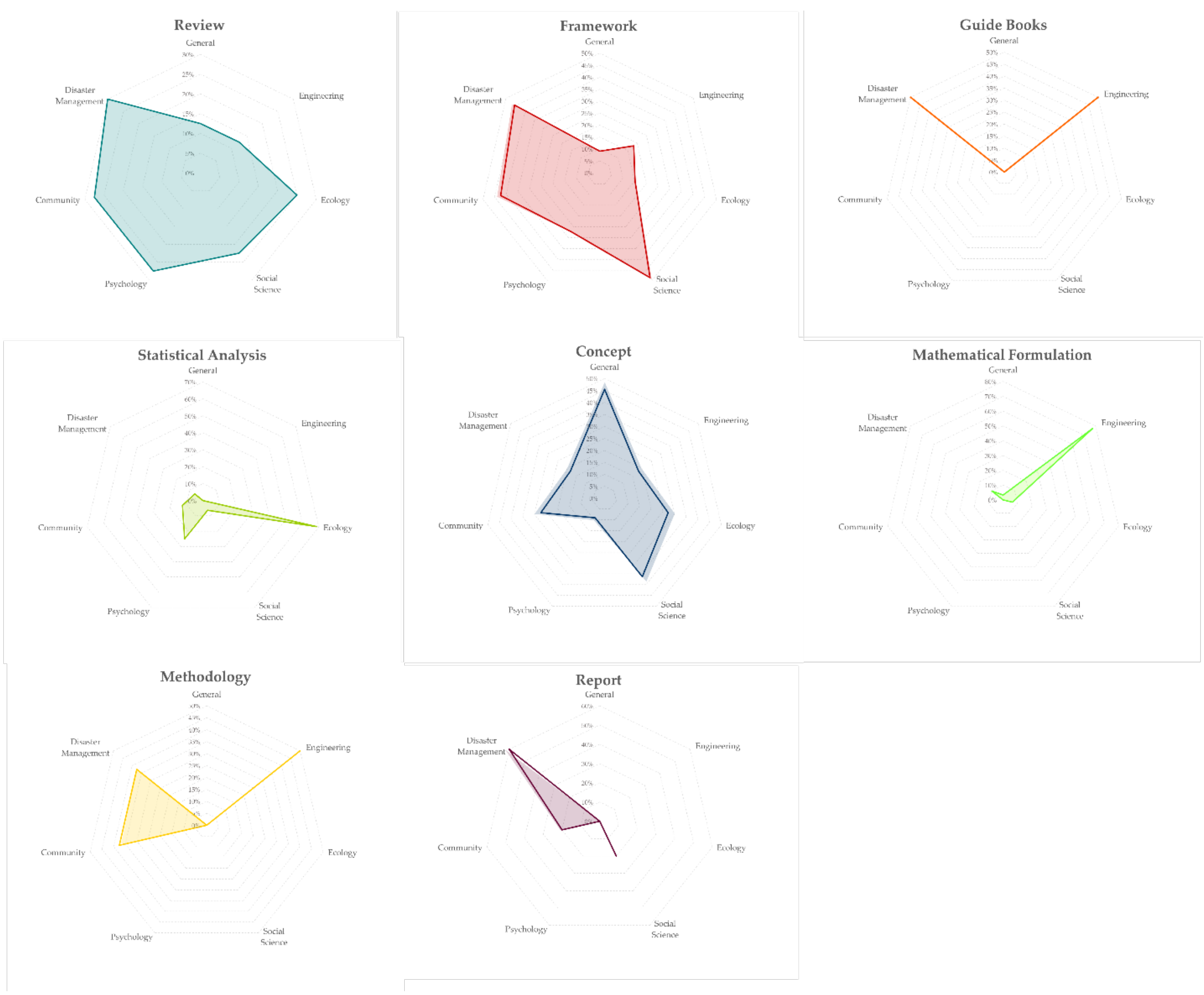

Figure 2. How Review, Framework, Guidebooks, Statistical Analysis, Concept, Mathematical Formulation, Methodology, and Report view the resilience through different disciplines

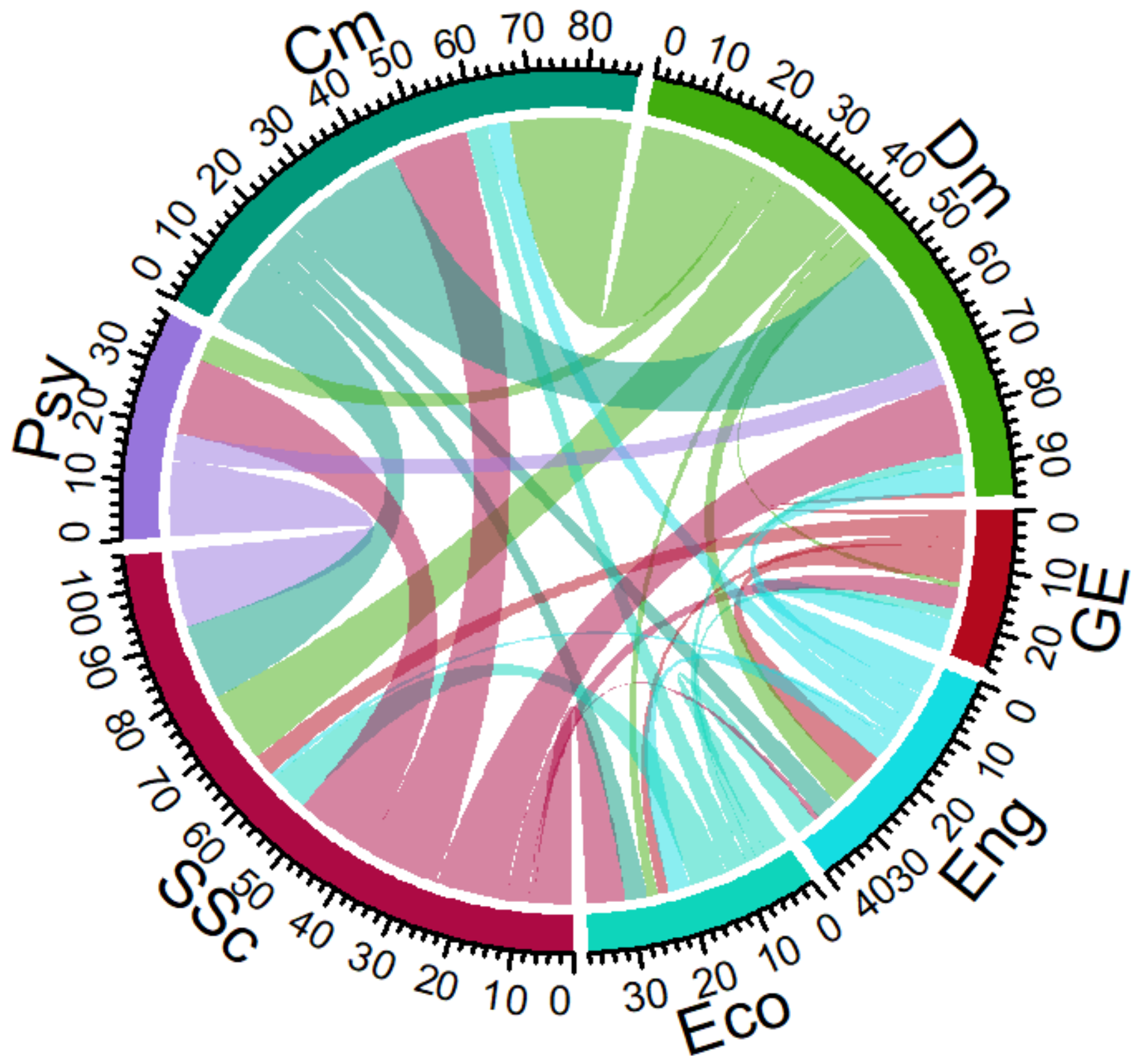

Figure 3. How papers in different disciplines integrate through the lens of General (GE), Ecology (Eco), Engineering (Eng), Psychology (Psych), Social Science (SSc), community (CM), and Disaster Management (DM), on resilience

Most articles fall under the ‘Quantification Model’ category, particularly within the Engineering discipline, which accounts for 28% of the total. This underscores the technical orientation of the field, where precise modeling and analytical methodologies are paramount. However, the ‘Conceptual Framework’ category also showcases a robust engagement with the Social Sciences and Disaster Management sectors, capturing 21% and 17%, respectively, which implies a recognition of the broader social and systemic implications of resilience. In addition, figure 3 employs a chord diagram to highlight the intricate relationships between these disciplines. In the chord diagram, each discipline is represented as a segment on the circumference of a circle, and the lines (chords) connecting these segments indicate relationships between them. The thickness of the chords represents the strength or amount of interaction between disciplines.

## 4. Resilience Concept

The term 'resilience' traces its roots back to the work of Young (1807), who described it as a property acting in opposition to impulses, much like strength counters pressure. He defined resilience as "The action which resists pressure is called strength, and that which resists impulse may properly be termed resilience. The resilience is jointly proportional to its strength and its toughness." This fundamental interpretation set the stage for a myriad of subsequent endeavors aiming to contextualize and deepen the understanding of resilience within various scientific disciplines.

Initially, in 1885, the term was used within the mechanical domain to depict steel beams' resistance and deformation behavior (Rankine, 1872). However, over time, the use of 'resilience' has expanded far beyond its original mechanical boundaries. Today, it is employed to evaluate systems' abilities to withstand and adapt to disruptions, spanning an array of multidisciplinary perspectives that include ecology (Gunderson et al., 2012), psychology (Levine, 2003), engineering (Hollnagel et al., 2006), and societal responses to significant environmental perturbations (Etingoff, 2016).

The Ball and Cup model serves as a visual metaphor in which a system's state is symbolized by a ball situated within a valley shaped like a cup. The cup's depth and width symbolize the system's resilience, indicating its ability to absorb disturbances, while the position of the ball within the cup reflects the system's vulnerability to changes. A myriad of external factors, such as sensitivity to environmental alterations, resource availability fluctuations, changes in physical or chemical properties, human activities (e.g., land use changes, pollution), and natural disasters (e.g., floods, fires), can exert forces on the ball. Meanwhile, adaptive capacity is represented by the ability of the system to modify the shape of the cup to fortify itself against disturbances. Resilience is intimately connected to the ideas of regime shifts and hysteresis. When disturbances compromise a system's resilience, it may experience a regime shift, transitioning from one stable state to an alternative one (Scheffer & Carpenter, 2003). The terminology associated with resilience within the Ball and Cup model is depicted in Table 1. Post-transition, hysteresis may manifest, which implies that reverting the conditions responsible for the shift does not necessarily restore the system to its initial state due to the emergence of new stabilizing feedback mechanisms and altered resilience and dynamics (Cross et al., 2009; Martin, 2012; Scheffer et al., 2001). While hysteresis elucidates the historical dependency and potential for multiple stable states within a system, transformability pertains to the system's capacity to undergo fundamental changes in state, often necessitated by surpassing the thresholds depicted in hysteresis loops.

Table 1: Summary of terminology needed linking to resilience

| Term | Definition | Ball and cup model |
|---|---|---|
| Hysteresis | When the path out is not the same as the path in or when it is difficult to reverse a regime change (Sguotti & Cormon, 2018) | |
| Regime shift/change | Persistent change in the structure, function, and mutual shift/change reinforced processes or feedback of an ecosystem. (Crawford, 1991) | |
| Transformability | When the existing system becomes untenable due to extreme changes in conditions, it needs to transform itself (point 1 in the ball and cup model) into a fundamentally new system with different structures, functions, and feedback (point 2 in the ball and cup model). (Resilience = evolve with the system) (Pinheiro et al., 2022; Walker et al., 2004) | |
| Absorptive | Ability to return rapidly and efficiently to the original state and **bounce back** (Resilience = return to earlier stability) | |
| Adaptive | Adaptive is a key aspect of how a system manages to stay resilient. It refers to the ability of the system's actors to manage and influence resilience by adjusting their behavior or by changing the rules or structure of the system to better cope with external changes or disturbances. Therefore, a highly adaptable system is typically also highly resilient. (Resilience = stay away from the threshold) | |

The complex interrelationships among resilience, adaptive capacity, and vulnerability form the cornerstone of numerous scholarly discourses in this field. However, a definitive understanding of these concepts remains an ongoing endeavor due to complex interrelationships among resilience, adaptive capacity, and vulnerability, as depicted in Fig. 4. Different researchers propose varied interpretations, with resilience often considered a key aspect of adaptive capacity. Yet, the focus and conceptualizations diverge based on specific theoretical orientations. Adaptive capacity is conceptualized by researchers such as (Adger, 2000a; Birkmann & Pelling, 2006; Folke, 2006) as the system's proficiency in managing change. This involves effectively leveraging resources like social networks, education, and technology to mitigate potential harm or to exploit emerging opportunities (Fig. 2.4-a1, a2). Conversely, vulnerability encapsulates the system's susceptibility to the adverse effects of change, often exacerbated by socioeconomic factors like poverty and social disparity, as highlighted by Manyena, (2006) and Cutter et al., (2008) (Fig. 4-b1, b2).

Resilience, as put forward by Bruneau et al., (2003); Paton & Johnston, (2017); Tierney & Bruneau, (2007), underscores the system's ability to endure disturbances and recover effectively (Fig. 4-a1, a2). Absorptive capacity, as defined by Gallopín, (2006) and Turner et al., (2003) (Fig. 4-c1), (Engle, 2011) (Fig. 4-c2), (Yoon et al., 2016) (Fig. 4-c3), emphasizes the system's capacity to assimilate new information or adopt innovative practices in a way that facilitates transformative processes. The interconnections among resilience, adaptive capacity, and vulnerability offer a vital perspective for comprehending system dynamics amidst change, deepening our understanding of the capacities and vulnerabilities that underlie system resilience while considering temporal perspectives that range from short-term resilience against immediate disasters to long-term adaptive capacities addressing persistent challenges like climate change. For example, a systems-oriented perspective might focus on elements' interconnectedness, whereas a human-centric approach might spotlight individual and collective actions (Bruneau et al., 2003).

By virtue of individualism versus collectivism, personal experiences, cultural contexts, and the researchers' belief systems can influence the perception of resilience or vulnerability. The research purpose—whether it is academic understanding, policy formulation, or intervention design—shapes how these concepts are utilized and defined Tierney & Bruneau, (2007). Though these variations yield a spectrum of perspectives, enhancing the understanding of these intricate concepts, they also challenge establishing a universally accepted definition. Nevertheless, these perspectives are pivotal to designing robust, flexible, and context-sensitive strategies in disaster management and resilience building.

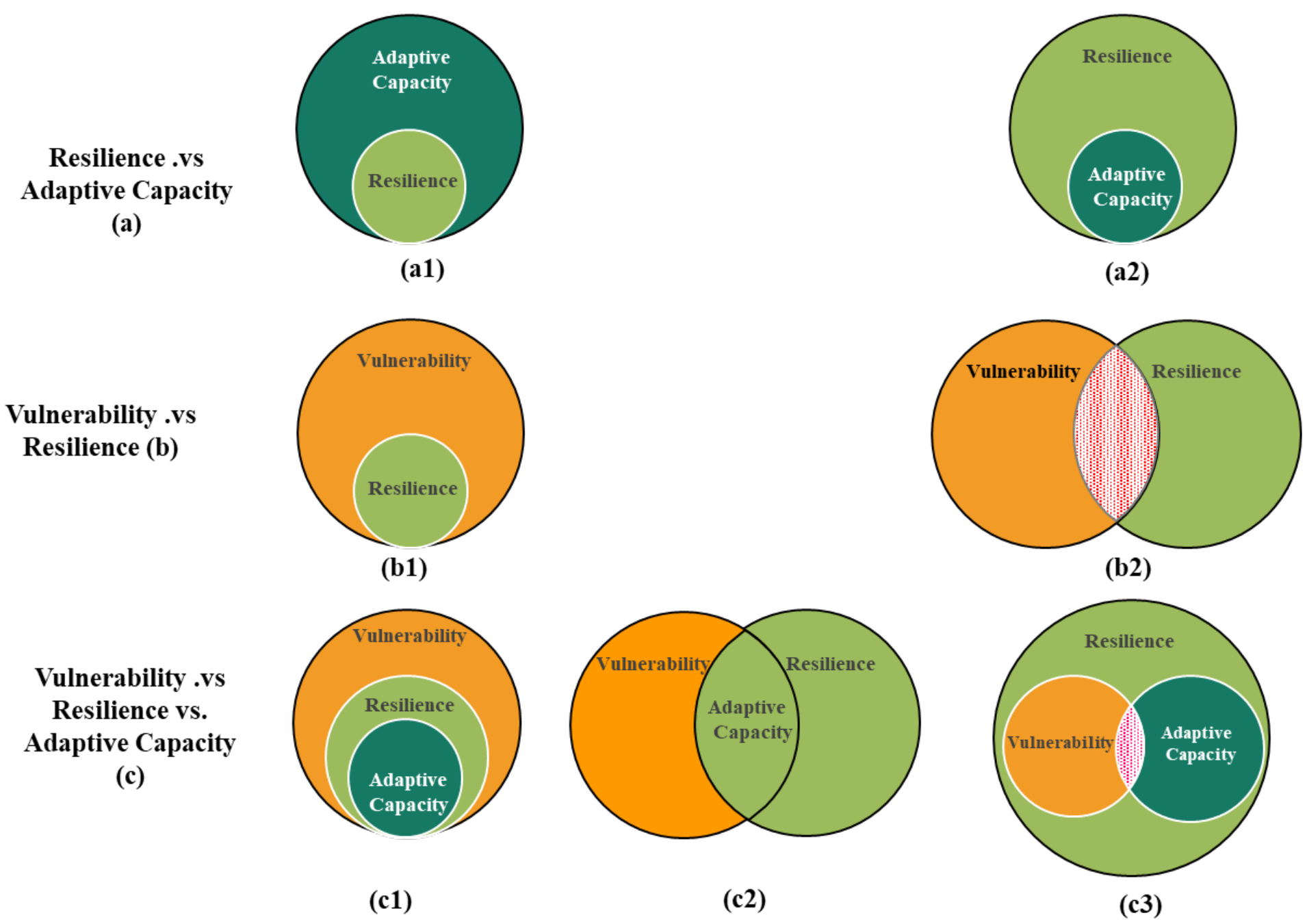

Figure 4. Conceptual linkages between vulnerability, resilience, and adaptive capacity (Cutter et al., 2008)

To gain a comprehensive understanding of the dynamics of a system, especially in terms of its resilience and adaptability, it is essential to measure various variables. These variables include “resistance”, which indicates how well the system can withstand disturbances; “elasticity”, showing its ability to regain its original state; “return time”, reflecting how quickly it recovers; “precariousness”, describing the inherent vulnerabilities; and “latitude”, representing the range within which the system can operate without undergoing a regime shift. These variables are further nuanced through the lens of panarchy theory, as developed by (Holling & Gunderson, 2002).

Panarchy introduces a multi-scalar, hierarchical structure of nested adaptive cycles that explores complex systems' cross-scale interactions and inherent dynamism. This intricate framework captures the dynamics across different scales. In panarchy's light, resilience is not just about steadiness or bouncing back but also encompasses a system's adaptability, ability to self-organize, and potential for transformation. Adaptive capacity, too, is viewed more expansively as not just reactive but also as the system's proactive evolution in the face of challenges or new possibilities (figure 5). Vulnerability is characterized as the susceptibility of a system to shocks, especially when it is reorganizing or undergoing major changes. Panarchy thus offers a more comprehensive lens for examining resilience, adaptive capacity, and vulnerability in unison Allen et al., (2014) For a detailed breakdown of resilience components, refer to table 2.

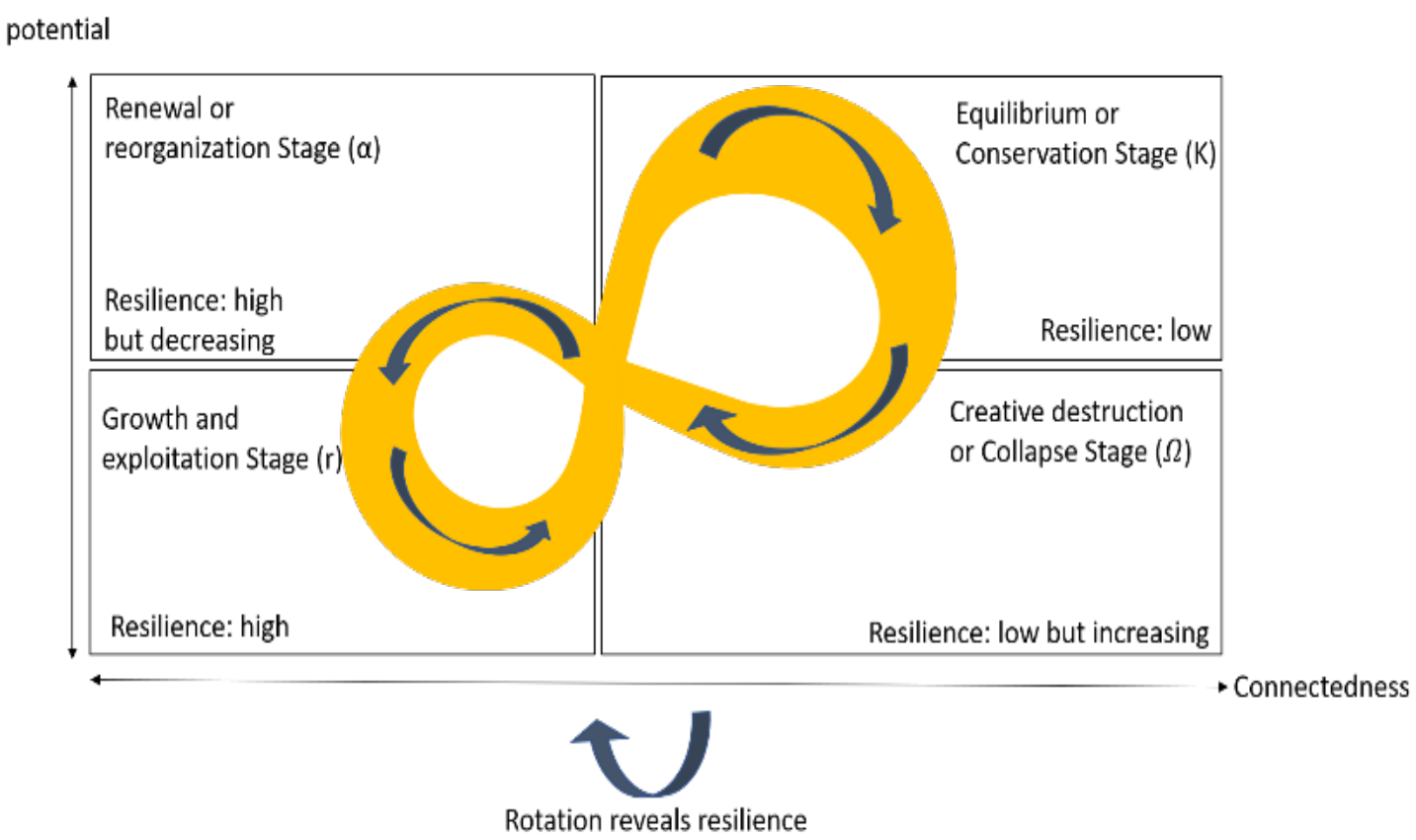

Figure 5. Adaptive Cycles (Holling, 2001)

Table 2: Resilience components (Walker et al., 2004)

| Term | Definition |
|---|---|
| Resilience | The resilience of a system is defined by its capacity to absorb disturbances, reorganize while undergoing change, and still retain essentially the same function, structure, identity, and feedback mechanisms. It has four components: latitude, resistance, precariousness, and panarchy. |
| Latitude | The maximum amount a system can change before it loses its ability to recover or crosses a threshold beyond which recovery becomes challenging or impossible. |
| Resistance | A measure of how difficult it is to change the system. The more resistant a system, the harder it is to alter its state. |
| Precariousness | A measure of how close the system's current state is to a threshold or limit that, if surpassed, would trigger significant change or disruption. |
| Panarchy | Considers the cross-scale interactions that influence resilience. The resilience of a system at a given scale depends on the states and dynamics at scales both above and below it. Factors like politics, market shifts, or climate change might include external influences. |

## 5. Resilience across disciplines

As we navigate the complexities of this broad-ranging concept of resilience, it is essential to consider different perspectives of resilience in disciplines. Some approaches towards quantifying resilience attempt to bridge the 'measurement gap' by identifying and implementing mathematical indicators and metrics distilled from academic literature. These measures are subsequently categorized utilizing the SMART (Specific, Measurable, Attainable, Relevant, and Time-based) criteria, which facilitate a structured and rigorous methodology for assessing resilience across a spectrum of applications. Conversely, other scholars opt for more qualitative approaches that emphasize the human and social dimensions of resilience. Some frameworks integrate quantitative and qualitative measures to offer a more holistic view of resilience. Thus, resilience is an encompassing concept that intersects with numerous disciplines; employing various methodologies and perspectives is vital. This diverse toolkit, ranging from SMART-based quantitative metrics to qualitative assessments and systems modeling, allows for a more nuanced understanding and application of resilience in tackling real-world challenges.

### 5.1. Resilience in Engineering

Engineering resilience is a concept that has been widely studied and implemented in various engineering disciplines. The term "engineering resilience" or "resilience engineering" is often used to describe the integration of resilience into engineering practices to design systems that can effectively respond to changes and recover from disruptions. Resilience engineering is maintaining elasticity when faced with stress (Thoma et al., 2016). From an engineering perspective, resilience is commonly discussed within the realm of physical sciences. Resilience engineering was proposed as an alternative or complement to the conventional view of safety; however, it has evolved into a dynamic approach to complex systems, focusing not on the reliability of individual components but rather on the active, anticipatory, and recovery capacities of the system as a whole, emphasizing the importance of understanding and facilitating real-world work processes, and challenging traditional concepts in safety research (Wears, 2006). In contrast to ecological resilience, which focuses on the system's capacity to absorb disturbances, prepare for future events, and integrate multiple resources, engineering resilience emphasizes the system's internal ability to rapidly reorganize and resolve disruptions without the need for external resources, indicating an inherent capability for long-term adaptation and learning from each event to adjust to internal and external changes (K.-H. Liao, 2012; Sharifi & Yamagata, 2014; Tri et al., 2017). Engineering resilience (as depicted in Figure 6) is the capacity of a system to return to a stable state following a disturbance. It emphasizes the safeguarding of the functional stability of engineering systems.

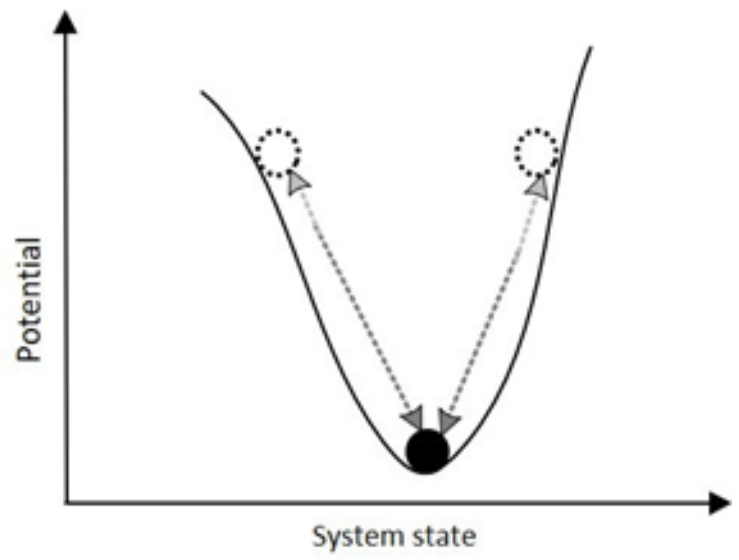

Figure 6. Engineering resilience (Tri et al., 2017)

In addition, resilience engineering is a field that focuses on the characteristics of resilient performance rather than resilience as property or quality or resilience in a dichotomy (*Hollnagel: What Is Resilience Engineering?*, 2022). Through questionnaires, measurable aspects of resilience such as adaptability, control, awareness, and time management have been discerned (Woods & Wreathall, 2011)

A comprehensive summary of reviewed articles contributes to a better understanding of the concept of engineering resilience in the engineering design community. Table 3 encompasses a diverse range of equations and formulas proposed by these articles to quantify and measure engineering resilience. These equations capture different aspects of resilience, such as the normalized area under the performance function (Ayyub, 2015; Rose, 2007), the ratio of performance loss (Zobel & Khansa, 2014), probabilistic events (Henry, 2012), state transition probabilities (Attoh-Okine et al., 2009; G. Cimellaro et al., 2008; G. P. Cimellaro et al., 2010; Y. Li & Lence, 2007; Munoz & Dunbar, 2015; Ouyang et al., 2012; D. Wang & Ip, 2009), and restoration of system performance (Ayyub, 2014; Cox et al., 2011; Dessavre et al., 2016; Dixit et al., 2016; Miller-Hooks et al., 2012; Ouyang et al., 2012; Ouyang & Duenas-Osorio, 2014; Pant et al., 2014).

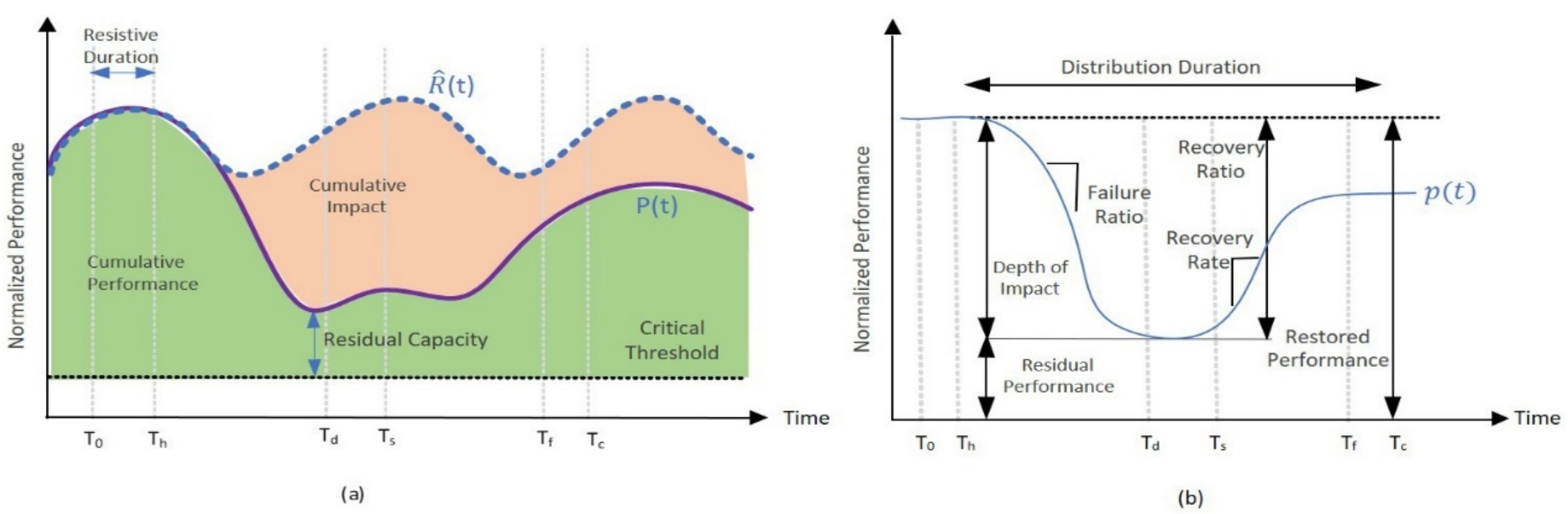

Figure 7. Engineering resilience (Tri et al., 2017) a) represents the resilience curve measured in actual units denoted as P(t), while b) represents the same curve normalized to a standard scale denoted as p(t).(Poulin & Kane, 2021)

The reviewed articles present a variety of resilience quantification metrics that have implications for designing and improving resilient-engineered systems. However, the diverse range of metrics also highlights the need for further research and development of universally applicable metrics, analysis methodologies, and design tools. These advancements would facilitate the quantification and enhancement of resilience across various engineering applications, providing valuable insights into the performance behavior of resilient systems. Figure 7 depicts an illustrative summary of some of the metrics for a resilience curve, presenting the system's adherence to the critical threshold while acknowledging the undefined disruptive duration within the control interval.

Table 3: Summary of Resilience in Engineering

| No. | Reference | Explanation | Formula |
| --- | --- | --- | --- |
| 1 | Bruneau et al., (2003) | Resilience Triangle<br>$Q(t)$: The system's infrastructure quality over time, from an initial to a restoration time. | $\psi_{\Delta}(t) = \int_{t_0}^{t_f} [100 - Q(t)]dt$ |
| 2 | Chang & Shinozuka, (2004) | Probabilistic resilience<br>$r^*$: Performance loss<br>$t^*$: Recovery time | $\psi = P(A \vert i) = P(r_0 < r^* \ and \ t_1 < t^*)$ |
| 3 | .Cox et al., (2011); Rose, (2007) | The Total System Economic Resilience $\psi TSER$ as the resilience to input supply disruptions, the difference between the maximum percent change in total output assuming a linear set of general equilibrium effects and the estimated percent change in total output considering a non-linear outcome.<br>$M$: The economy-wide input-output multiplier<br>$\%\Delta DY^m$: The maximum percent change in direct output<br>$\%\Delta Y$:: The estimated percent change in total output | $\psi TSER = (\%\Delta DY^m - \%\Delta Y)/(\%\Delta DY^m)$ |
| 4 | Y. Li & Lence, (2007) | Resilience is measured as the recovery probability from a system failure.<br>$\mathrm{g(t_1)}$ and $\mathrm{g(t_2)}$: The performance functions at $\mathrm{t_1}$and $\mathrm{t_2}$ respectively.<br>P: Probability<br>$P_F$ : The failure probability<br>$P_{FS}(\mathrm{t_1,t_2})$: The probability that the system fails at $\mathrm{t_1}$ and recovers at $\mathrm{t_2}$ | $\psi(\mathrm{t_1,t_2}) = P(\mathrm{g(t_2)} \geq 0 \vert \mathrm{g(t_1)} < 0)$<br>$= \frac{P(\mathrm{g(t_1)} < 0 \cap \mathrm{g(t_2)} \geq 0)}{\mathrm{P(g(t_1)} < 0)} = \frac{P_{FS}(\mathrm{t_1,t_2})}{P_{FS}(\mathrm{t_1})}$ |
| 5 | D. Wang & Ip, (2009) | Resilience is measured based on the reliability of each supplier in a networked system following a failure.<br>$\mathrm{n = n_1 + n_2}$: total number of two kinds of nodes<br>$\mathrm{m} = \vert E \vert$: Number of edges in the set<br>$d_i$ , $i = 1, \dots, n_1$: Demand for node<br>$s_j$ , $j = 1, \dots, n_2$: Available supply of node j ∈ S (The supply capacity in a unit of time)<br>$p_j$ , $j = 1, \dots, n_2$: Supply reliability of node<br>$c_k$ , $k = 1, \dots, m$: Flow capacity of edge (delivery line) $k \in E$<br>$q_k$ , $k = 1, \dots, m$: Reliability of edge | $\psi_i = \frac{p_j q_k \ min\{d_i, s_j, c_k\}}{d_i}$ |
| 6 | Renschler et al., (2010) | Resilience as the normalized area under the system response after a disruption.<br>Q(t): Functionality of the system as a function of time<br>$\mathrm{T_{LC}} = t_c - t_h$ : Control time, representing the period over which resilience is assessed (Figure 7: distributed duration) | $\psi = \int_{t_h}^{t_h + T_{LC}} \frac{Q(t)}{T_{LC}} dt$ |
| 7 | G. P. Cimellaro et al., (2010) | Resilience combines the integrals of system performance before and after a disruption.<br>$\mathrm{Q_1}(t)$ and $\mathrm{Q_2}(t)$: Normalized waiting time with a distinction made for waiting time before and after a critical condition<br>$\mathrm{WT_c}$: Waiting time at the maximum capacity of the hospital<br>$\mathrm{WT_0}$ : Waiting time in normal operational conditions pre-disaster | $Q(t) = \begin{cases} Q_1(t) = \frac{max(WT_c - WT_0) - WT_0}{WT_c - WT_0} & for \ WT < WT_c \\ Q_2(t) = \frac{WT_c - WT_0}{max(WT_c, WT - (WT_c - WT_0))} & for \ WT \geq WT_c \end{cases}$<br>$\psi(t) = \alpha \int_{t_h}^{t_c} \frac{Q_1(t)}{\mathrm{T_{LC}}} dt + (1 - \alpha) \int_{t_h}^{t_c} \frac{Q_2(t)}{\mathrm{T_{LC}}} dt$ |

| No. | Reference | Explanation | Formula |
|---|---|---|---|
| | | WT: Waiting time during the transient condition<br>$\alpha$: Weighting factor used to combine the integrals of the two functionalities<br>$T_{LC} = t_c - t_h$ | |

Table 3 (continued)

| No. | Reference | Explanation | Formula |
|---|---|---|---|
| 8 | Youn et al., (2011) | Resilience as the sum of passive survival rate and proactive survival rate. | $\psi = R(reliability) + \rho(restoration)$ |
| 9 | Zobel, (2011) | Extends the resilience triangle to scenarios of partial recovery from multiple disruptive events.<br>$X$: Initial loss<br>$T$ : Recovery time<br>$T^*$: long time interval | $\psi(X,T) = \frac{T^* - XT/2}{T^*} = 1 - \frac{XT}{2T^*}$ |
| 10 | Henry, (2012) | Resilience is the resistance measure of a system to disruption during different stages.<br>$\psi(t_r \mid e_j)$: is the figure-of-merit evaluated at time $t_r$ under disruptive event $e_j$<br>$t_r$: Time of evaluation, and $t_r \in (t_d, t_f)$<br>$t_0$: Figure-of-merit in the original stable state<br>$\psi(t_d \mid e_j)$: Figure-of-merit in the disrupted state | $\psi(t_r, e_j) = \frac{\psi(t_r \mid e_j) - \psi(t_d \mid e_j)}{\psi(t_0) - \psi(t_d \mid e_j)}$ |
| 11 | Dixit et al., (2016); Miller-Hooks et al., (2012) | Resilience is the expected fraction of demand that can be satisfied post-disaster.<br>$D_w$: The original pre-disaster demand for O–D pair network $w$<br>$d_w$: The original pre-disaster demand for O–D pair network $w$ | $\psi = E[\frac{\sum_{w \in W} d_w}{\sum_{w \in W} D_w}]$ |
| 12 | Ouyang et al., (2012); Ouyang & Duenas-Osorio, (2014) | Resilience as normalized performance loss due to multiple disruptive events.<br>$\psi$: Expected Annual Resilience; a metric evaluating system resilience over a year by comparing actual to target performance.<br>$T$ : Time interval for a year<br>$P(t)$: Actual performance curve as a function of time<br>$TP(t)$: Target performance curve as a function of time, $t$<br>$n$: Event occurrence number<br>$N(T)$: Total number of event occurrences during time $T$.<br>$t_n$ : Occurrence time of the nth event; a random variable.<br>$AIA_n(t_n)$: Impact area for $n^{th}$ event at time $t_n$ | Environmental Resiliency:<br>$\psi = \frac{\int_{t_0}^{T} TP(t)dt - \sum_{n=1}^{N(T)} AIA_n(t_n)}{\int_{t_0}^{T} TP(t)dt}$ |
| 13 | Omer et al., (2013) | Environmental resiliency and Cost resiliency in the case of transportation networks | Environmental Resiliency:<br>$\psi = R_{ENV} = \frac{Env_Impact_{before_shock}}{Env_Impact_{after_shock}}$<br>Cost Resiliency:<br>$\psi = R_{Cost} = \frac{Cost_{before_shock}}{Cost_{after_shock}}$ |
| 14 | Ouyang & Duenas-Osorio, (2014) | Resilience as one minus expected loss, assuming a constant baseline and Poisson-based disruptive events.<br>$\lambda$: The occurrence rate of the hazards per year | $\psi = 1 - \lambda E[IA]$ |

| | | $E[IA]$: The expected impact area under the hazard, accounting for all possible hazard intensities | |
|---|---|---|---|
| 15 | Shafieezadeh & Burden, (2014) | Resilience as the ratio of post-disruption to baseline system response.<br>$P_R(t)$: Actual system performance over time considering disruptions and restorations.<br>$P_T(t)$: Target performance curve; desired or expected system performance over time. | $\psi = \frac{\int_{t_0}^{T} P_R(t)dt}{\int_{t_0}^{T} P_T(t)dt}$ |

Table 3 (continued)

| No. | Reference | Explanation | Formula |
|---|---|---|---|
| 16 | Francis & Bekera, (2014) | Incorporates recovery speed in resilience measurement.<br>$S_p$ : Speed recovery factor.<br>$F_0$ : Original stable system performance level.<br>$F_d$ : Performance level immediately post-disruption.<br>$F_r$: Performance at a new stable level after recovery efforts have been exhausted.<br>$t_\delta$ : Slack time is the maximum amount of time post-disaster that is acceptable before recovery begins.<br>$t_r$ : Time to final recovery (i.e., new equilibrium state).<br>$t_r^n$ : Time to complete initial recovery actions.<br>$a$: Parameter controlling decay in resilience attributable to time to a new equilibrium. | $\psi_i(S_p, F_r, F_d, F_0) = S_p \frac{F_r}{F_0} \frac{F_d}{F_0}$<br>$S_p = \begin{cases} \frac{t_\delta}{t_r^n} exp[-a(t_r - t_r^n)] & for\ t_r \geq t_r^n, \\ \frac{t\delta}{t_r^n} & otherwise. \end{cases}$ |
| 17 | Ayyub, (2014) | Incorporates the effect of system aging in the resilience assessment.<br>$T_i$: The initial time before the failure<br>$\Delta T_f$ : The duration of the failure event (f)<br>$\Delta T_r$ : The duration of the recovery event (r)<br>$F$ and $R$: The system's performance during failure and recovery, respectively | $\psi = \frac{T_i + F\Delta T_f + R\Delta T_r}{T_i + \Delta T_f + \Delta T_r}$ |
| 18 | Zobel & Khansa, (2014) | Resilience as the normalized area of the performance loss triangle, assuming linear recovery.<br>$\psi_i$ : The partial resilience associated with sub-event $i$<br>$X_i$ and $\acute{X}_i$: The decline and recovery in performance due to the sub-event<br>$T_i$ : The time until the next sub-event occurs or the time for the system to recover from the sub-event<br>$T_n$ : The maximum allowable time for recovery | $\psi_i = 1 - \frac{(X_i + \acute{X}_i).T_i}{2T_n}$ |
| 19 | Munoz & Dunbar, (2015) | Resilience of supply chain resilience by considering the weighted sum of various dimensions.<br>$R_m$ and $R_r$: Resilience scores for the manufacturer and retailer respectively<br>$w_{i,m}$ and $w_{i,r}$: The standardized weight factors for each dimension of resilience across the manufacturer and retailer supply chain tiers, respectively. R: Recovery<br>$PL$: Performance Loss<br>$PrL$: Profi le Length<br>$WS$: Weighted Sum<br>I: Impact | $\psi_m = w_{1,m}.R_m + w_{2,m}.I_m + w_{3,m}.PL_m + w_{4,m}.PrL_m + w_{5,m}.WS_m$<br>$\psi_r = w_{1,r}.R_r + w_{2,r}.I_r + w_{3,r}.PL_r + w_{4,r}.PrL_r + w_{5,r}.WS_r$<br>$\psi = \psi_m + \psi_r$ |
| 20 | Bhavathrathan & Patil, (2015) | Resilience as the normalized difference between optimal and actual capacity. | $\psi = \frac{STT^* - (SO - STT)}{STT^*}$ |
| | | $STT^*$: The critical operational cost<br>$SO - STT$: The best operational cost or system optimal travel time | |
| 21 | Ayyub, (2015); Woods & Wreathall, (2011) | Resilience as practicality desired for systems with time invariant performance<br>$\lambda$: Rate of stressor s, i.e., Poisson process es where the planning horizon (t) is equal to the return period ($\frac{1}{\lambda}$)<br>$p$: Probability of failure given a stressor, i.e., inherent strength of the system<br>$Q_{100}$ : Capacity of the system<br>$Q_r$ : Robustness of the system<br>$t$: Planning horizon $t$ | $\psi = 1 - exp[-\lambda t(1 - p\bar{R}_f)] + exp(-\lambda t)$<br>$\bar{R}_f = \frac{(t_r - t_i)(Q_{100} - Q_r)}{Q_{100} t}$ |

Table 3 (continued)

| No. | Reference | Explanation | Formula |
|---|---|---|---|
| 22 | Yodo & Wang, (2016a, 2016b) | Resilience as a function of disruptions, system-specific characteristics, reliability, and restoration through conditional probabilities. | $\psi$<br>$= P(Disruptions)$<br>$\times P(Characteristics\ Events \mid Disruptions)$<br>$\times P(Reliability \mid SystemCharacteristics)$<br>$\times P(Restoration\ \mid Reliability, System\ Characte$<br>$\times P(Resilience\ \mid Restoration, Reliability, Syster$ |
| 23 | T.-Y. Liao et al., (2018) | Resilience as the expected mean of the ratio of the area between the actual performance curve and timeline to the area between the target performance curve and timeline over a given period of time.<br>$AP(t)$: Actual performance function curve<br>$TP(t)$: Target performance function curve<br>$T$: given time interval from the disaster occurrence to the complete restoration of the system performance. | $\psi = E[\int_0^T \frac{AP(t)}{TP(t)} dt]$ |
| 24 | Twumasi-Boakye & Sobanjo, (2018) | Improved model accounting for recovery time and state for damaged infrastructure restoration.<br>$T_{mod}$: Time for restoring moderately damaged infrastructure in days<br>$T_{ext}$ : Time for restoring extensively damaged infrastructure in days<br>$T_{comp}$: Time for restoring completely damaged infrastructure in days | $\psi(t_f \mid e) = 1 - \frac{1}{\bar{\bar{T}}}[\int_0^{T_{mod}} [1 - \psi_k^{p(h)}(t)]$<br>$+ [\int_{mod}^{T_{ext}} [1 - \psi_k^{p(h)}(t)]]$<br>$+ [\int_{T_{ext}}^{T_{comp}} [1 - \psi_k^{p(h)}(t)]]$ |
| 25 | Zhang et al., (2019) | Resilience as the total travel time.<br>$\psi(t_f \mid e)$: system performance over time<br>$t_f$ , $t_0$ , and $t_d$: different time points | $\psi(t_f \mid e) = 1 - \frac{\psi(t_f \mid e) - \psi(t_0)}{\psi(t_d \mid e) - \psi(t_0)}$<br>$= \frac{\psi(t_d \mid e) - \psi(t_f \mid e)}{\psi(t_d \mid e) - \psi(t_0)}$ |
| 26 | Zhang et al., 2019) | Traffic resilience defined based on spatiotemporal jammed 3D cluster of size S which is the loss of resilience in the traffic network<br>$M_{\psi S}(t)$: jammed cluster size over time<br>$T$ : The recovery duration<br>$\psi S$: The cluster size, the loss of resilience in the traffic network | $\psi S = \int_{t_0}^{t_c} M_{\psi S}(t) dt$ |
| 27 | Watson et al., (2021) | Resilience as SSoSRM combines elements of different approaches to quantify resilience for systems-of-systems. Combined (Bruneau et al., 2003; Zobel, 2011) | $\psi_{SoSSRM} = \frac{\int_{t_0}^{t_0+T^*} Q(t)\, dt}{T^*. Q(0)}$ |

| $Q(t)$: the system's measure of performance<br>$Q(0)$: Initial performance |
| --- |

## 5.2. Resilience in Ecology

The concept of resilience in ecological systems first gained traction in the 1970s, notably through the work of Holling, (1973), who introduced it as one of two defining properties of such systems, the other being stability. Resilience was viewed as an inherent quality of the system, with stability defined as its ability to revert to an equilibrium state following a significant disruption to its structural integrity. Holling, (1973) described resilience as measuring the persistence of relationships within a system and its capacity to absorb shifts in state variables, driving variables, and parameters while maintaining its overall functionality. This concept was visualized through the ball-and-cup model, where the depth of the 'cup' signifies the system's resilience, representing its capacity to withstand disturbances before a state change occurs. Figure 8 shows the dynamics of the system. Deep cups symbolize high resilience, while shallow ones indicate vulnerability to change, thus illustrating the system's recovery potential from disturbances, essentially defining ecological resilience.

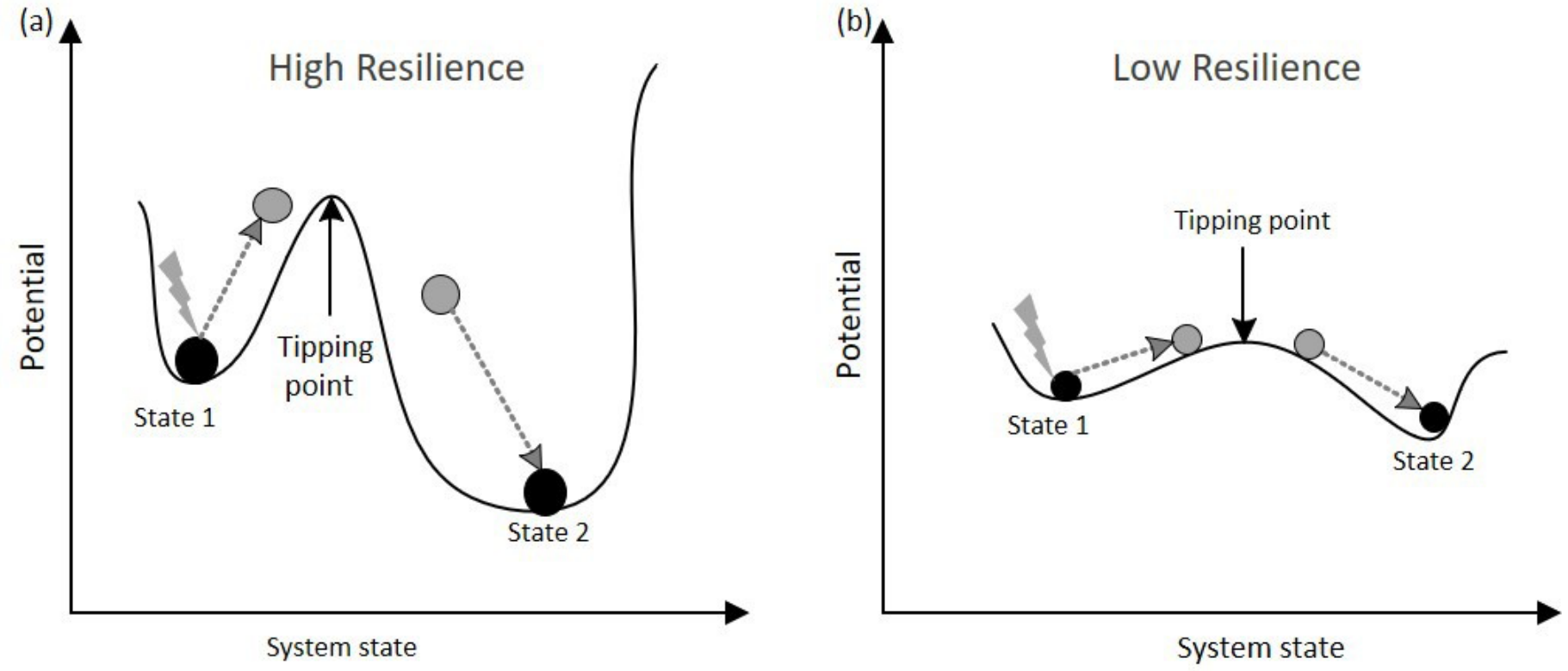

Figure 8. Ecological resilience (Holling, 1973)

Within the context of ecology, definitions of ecological resilience presuppose that a system can exist in multiple alternative states resistant to change, i.e., a system can absorb a considerable amount of disturbance before transitioning from one state to another (Carpenter & Walker, 2001; Perrings, 1998; Pimm, 1984). It's worth noting that disturbances can be artificially induced, encompassing conditions such as heat, fire, drought, and forests, which are especially relevant in studies focusing on soil and forest resilience (figure 9 and figure 10) (Y. Li & Lence, 2007; Lloret et al., 2011; Matos et al., 2020; Nimmo et al., 2015; Yi & Jackson, 2021). This interpretation significantly differs from engineering resilience, which primarily focuses on the system's resistance to change and assumes a singular stability regime (Holling, 1973; Zampieri, 2021). Contemporary perspectives on ecological resilience encompass various components, including resistance and recovery (Gunderson, 2000; Hodgson et al., 2015; Ingrisch & Bahn, 2018). Table 4 summarizes the resilience measurement in the context of ecology and resilience.

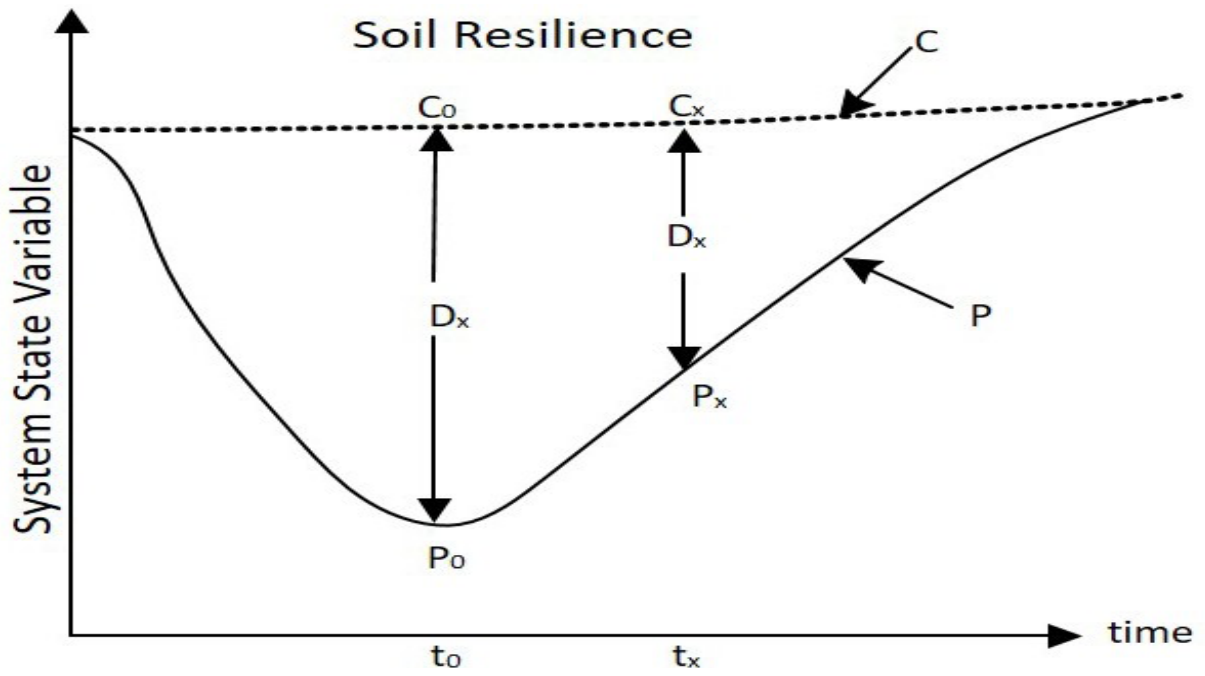

$$\text{Resilience} = \frac{P_x}{C_0}$$

Figure 9. Soil- ecological resilience. Here, $D_0 = C_0 - P_0$ is the difference between the control ($C_0$) and the disturbed soil ($P_0$) at the end of the disturbance ($t_0$). (Biggs et al., 1999; Kaufman, 1982; Sousa, 1980)

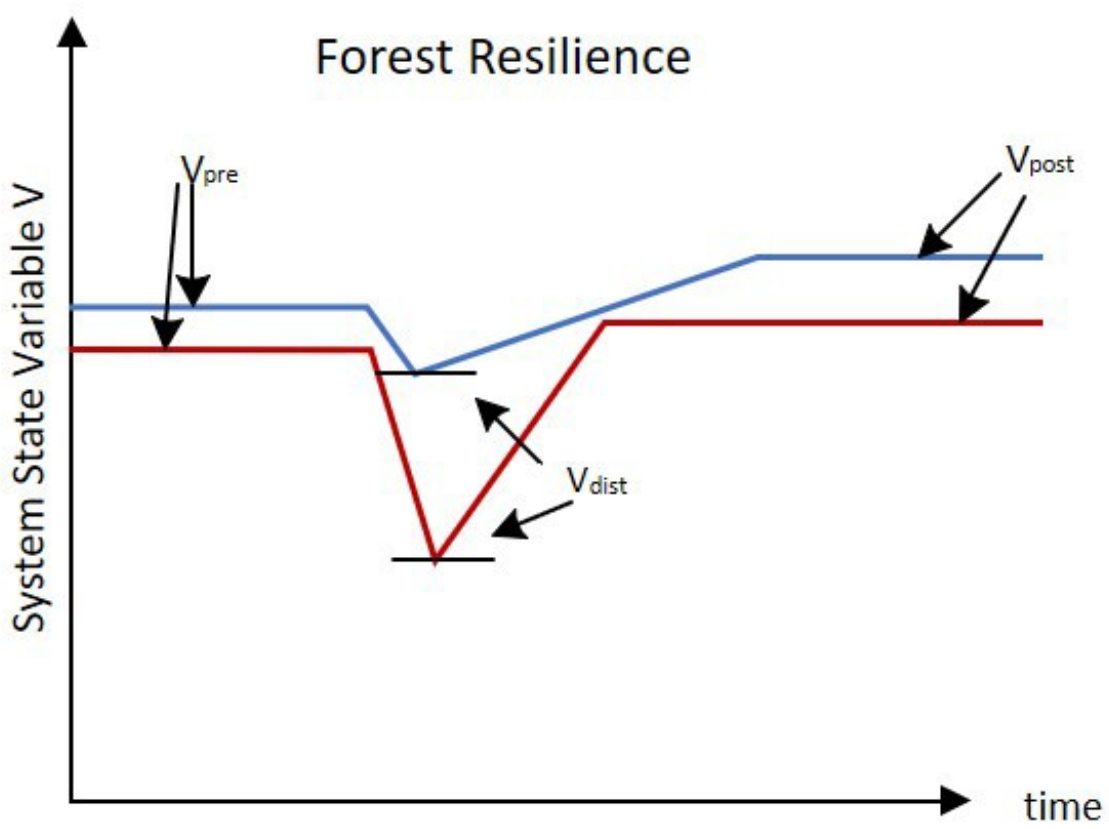

Figure 10. Forest- ecological resilience- $v_{pre}$ is the average value of a state variable in years preceding the disturbance, and $V_{post}$ is the average value of post-disturbance years. The averaging period used by different investigators ranged from 1 year to 11 years. (Lloret et al., 2011; Matos et al., 2020; Nimmo et al., 2015)

Table 4. Summary of resilience in ecology articles

| No. | Reference | Explanation | Formula |
|---|---|---|---|
| 1 | Biggs et al., (1999); Kaufman, (1982); Sousa, (1980) | Introduces a formula for soil resilience in ecological systems, using the difference between control and disturbed soil after a disturbance. | $\text{Resilience} = \frac{P_x}{C_0}$ (figure 9) |
| 2 | Lloret et al., (2011); Matos et al., (2020); Nimmo et al., (2015) | Introduces formulas for forest resilience in ecological systems, using the average value of a state variable before and after a disturbance. | $\text{Resilience} = \frac{V_{post}}{v_{pre}}$<br>$\text{Resilience} = \frac{V_{post}-v_{pre}}{v_{pre}}$<br>(figure 10) |
| 3 | Arnoldi et al., (2018); Guillot et al., (2019); Neubert & Caswell, (1997); Orwin et al., 2006; Rivest et al., (2015); Woods & Wreathall, (2011), (2011); Yeung & Richardson, (2016) | Introduces a dimensionless resilience metric that allows for comparison across diverse fields. This metric accounts for both short-term and long-term resilience. | $\text{Resilience} = \frac{2\times\lvert S_R-S_o\rvert}{\lvert S_R-S_o\rvert+\lvert S_R-S_Y\rvert} - 1$ |
| 4 | Scheffer et al., (2001) | General ecology and proposes a measure for resilience as the distance a system can be perturbed before it shifts into a different state. | $\text{Resilience} = \lvert E - E_c\rvert$ |
| 5 | Cantarello et al., (2017) | Forest ecology and introduces a measure for resilience in the context of forest ecosystems based on tree-ring data. | $\text{Resilience} = 1 - \frac{A_{disturbance}}{A_{Control}}$ |
| 6 | Oliver et al., (2015) | Forest ecology and proposes a measure for resilience based on the recovery of tree growth after | $\text{Resilience} = \frac{G_{recovery}}{G_{pre-disturbance}}$ |

**Nomenclature:**

| | |
|---|---|
| $S_R$: | Reference state |
| $S_o$: | State at the detection of maximum deviation |
| $S_Y$: | State at a present time point after the detection of maximum deviation. |
| E: | The current state of the system |
| $E_c$: | Critical threshold |
| $A_{disturbance}$: | Area affected by the disturbance |
| $A_{control}$: | Control area unaffected by the disturbance |
| $G_{recovery}$: | Tree growth during the recovery period |
| $G_{pre-disturbance}$: | Tree growth before disturbance |

### 5.3. Resilience in Psychology

Resilience has been a concept in psychology for a longer period than in ecology. (Garmezy, 1971). The term resilience started to be used in psychology in the 1950s; however, in the late 1980s, it became more popular (Flach, 1988), particularly concerning the psychiatric problems of children (Goldstein & Brooks, 2005). Resilience in Psychology concentrates on how individuals handle the external world's influences, while engineering prioritizes the durability and endurance of structures before and after catastrophic events (Wisner & Kelman, 2015).

In psychology, resilience is a dynamic process of positive adaptation within the context of significant adversity such as trauma (Hernández et al., 2007), tragedy (Stow, 2017), threats (Seery, 2011), or significant sources of stress (Luthar et al., 2000). Resilience, or the ability to bounce back from adversity, is increasingly recognized as an essential factor in mental health. Various studies have identified protective elements that support resilience, including hardiness (Bonanno & Mancini, 2008), positive emotions (Tugade & Fredrickson, 2004), extraversion (Campbell-Sills et al., 2006), self-efficacy (Gu & Day, 2007), self-esteem (Kidd & Shahar, 2008). Resilience is characterized as the process and outcome of adeptly adapting to challenging life circumstances, influenced significantly by an individual's worldview, the presence and quality of their social support, and their coping strategies (*American Psychological Association*, 2016). Furthermore, psychological research suggests that this ability for superior adaptation, or heightened resilience, can be nurtured and refined. Rodriguez-Llanes et al., (2013) reviewed six empirical studies that focus exclusively on disaster settings to identify 53 indicators of psychological resilience, including social support, gender, coping strategies, self-esteem, problem-focused coping, emotion-focused coping, professionalism, self-efficacy, motivation to teach, commitment to school and profession, capacity to cope with difficult conditions, talent in behavior management, sense of pride and satisfaction in enacting job, and posttraumatic growth, which the most consistent indicators of psychological resilience were social support and gender (Graber et al., 2015).

Resilience research draws insights from various related fields. This multidisciplinary approach is depicted in the visualization of resilience research, highlighting its intersections with key disciplines within psychology. A comprehensive analysis of psychological resilience has identified several critical dimensions of resilience. These dimensions are outlined in Table 5 below, which summarizes the psychological branches instrumental in building resilience.

Table 5. Summary of branches of Psychology for building resilience

| **Branch of Psychology** | **Indicators of Resilience** | **Reference** |
|---|---|---|
| Developmental | Children's responses to adversity: Effective | (Clauss-Ehlers, 2008) |
| Psychology | Coping, Positive emotions, Flexible use of emotional resources, Adaptive capacities | (Masten & Wright, 2010) |
| Traumatology | Adult responses to trauma: Cognitive reappraisals related to cognitive therapy, Capacity to recover from negative events and stress inoculation, Control over the process of remembering traumatic experiences | (Salloum & Lewis, 2010) |
| Humanistic Psychology | Human meaning-making and growth: Self-esteem, Personal autonomy, Meaning of one's life, Empathy | (Bonanno, 2008; Bonanno & Mancini, 2008; Bonnano, 2004; Masten & Obradovic, 2008) |
| Health Psychology | Active coping style in confronting a stressor, including exercise and training. Positive emotions, including optimism and humor, the Capacity to convert traumatic helplessness into learned helpfulness | (Almedom & Glandon, 2007; Netuveli et al., 2008; Rodriguez-Llanes et al., 2013) |
| Neuro-biological Psychology | Psychosocial protective mechanisms with genetic, epigenetic, and physiological processes | (Feder et al., 2009; Wu et al., 2013) |

### 5.4. Resilience in Social Science

Resilience studies gained momentum in the social sciences towards the end of the 1990s (Batabyal, 1998). Fundamentally, social resilience relates to the competencies of various social units, including individuals, organizations, or entire communities. It emphasizes their ability to endure, absorb, manage, and adapt to a broad spectrum of environmental and societal risks they might encounter (Keck & Sakdapolrak, 2013). The various definitions of social resilience have led to numerous social resilience frameworks. Each framework brings its unique strengths and weaknesses to the fore, and several categories can sometimes overlap. (Saja et al., 2019) evaluated 31 frameworks into four classifications: conceptual foundations; Coping, Adaptive, and Transformative/Participative capacities (CAT capacities); Social and Interconnected Community Resilience dimensions; and Structural and Cognitive dimensions of social resilience (Table 6).

Table 6. Descriptions of Social Resilience Framework Classifications

| No. | Classification | Description |
|---|---|---|
| 1 | Capital-based dimensions | Frameworks centered around social capital's role in resilience, but often incomplete. |
| 2 | Coping, Adaptive, and Transformative Capacities (CAT Capacities) | Frameworks focus on communities' ability to manage and adapt to disasters yet can be subject to varied interpretations. |
| 3 | Social and interconnected community resilience dimensions | Frameworks that incorporate social resilience as a part of broader community resilience often lack specific details. |
| 4 | Structural and cognitive dimensions of social resilience | Frameworks that divide social resilience into tangible (structural) and intangible (cognitive) factors but require further categorization for practical use. |

The concept of dimensions of social science resilience is related to the ability of human communities to withstand and recover from shocks, stresses, and adversities (Table 7) (Frankenberger et al., 2013).

Table 7. Dimensions of social science resilience

| No. | Dimension | Description | Reference |
|---|---|---|---|
| 1 | Preparedness | The collective knowledge and memory of a community aids in preparing for potential challenges or disasters. | (Jabeen et al., 2010) |
| 2 | Responsiveness | The ability of a community to effectively act during a crisis is supported by strong social structures and networks. | (*World Economic Forum*, 2013) |
| 3 | Learning and Innovation | Learning and Innovation | (Marshall & Marshall, 2007) |
| 4 | Self-Organization | Communities' ability to organize themselves, especially in times of crisis. This includes elements of human capital, motivation, and leadership. | (*World Economic Forum*, 2013) |
| 5 | Diversity | Involves a wide range of resources, ideas, and solutions, including social diversity, to strengthen a community's resilience. | (Berkes, 2007) |
| 6 | Inclusion | Represents the active involvement of diverse community members in decision-making and planning processes. Power dynamics and representation play key roles here. | (Béné et al., 2012; Davidson, 2010; Leach, 2008) |
| 7 | Aspirations | Reflects the future-oriented vision of community members.<br>Influences the investments made towards improving community well-being | (Agency (FEMA), 2019) |

Strong bonds within close-knit communities, being able to depend on others during difficult times, and having open communication between different groups are typically indications of a healthy social connection. Social capital is crucial in enhancing community resilience by offering support to those impacted by disasters, facilitating coordinated local responses to adapt to difficulties, and empowering the community to bring about positive changes through a united voice. Bonding, Bridging, and Linking Social Capital are social capital forms mutually reinforcing resources in a disaster. Addressing the forms of social capital involves delving into intricate questions about its nature and the ways different forms interact to produce specific adaptive outcomes. Researchers typically categorize social capital into bonding (relationships with family and relatives), bridging (relationships with neighbors and friends), and linking (relationships between individuals, communities, and organizations structured differently in societal power) (Azad & Pritchard, 2023; Islam & Walkerden, 2015; Saha & others, 2021). Besides that, the social vulnerability index was created to assist in guaranteeing the safety and welfare of their population. The Social Vulnerability Index (SVI) (*Centers for Disease Control and Prevention (CDC)*, 2022) is a critical tool for assessing community resilience in the face of public health threats, whether they emerge from disease outbreaks or natural and human-caused emergencies. This index incorporates 16 distinct vulnerability measures across socioeconomic, demographic, and housing/transportation themes. By evaluating census tracts against these measures—particularly highlighting those in the top 10% of vulnerability with flags—the SVI facilitates the identification of populations that may require additional support. Furthermore, it's essential to approach the SVI with an understanding of its initial screening role, acknowledging that it may necessitate further situation-specific analyses and incorporating more localized data for greater accuracy. The SVI is grounded in data from the American Community Survey and is adapted from a model initially developed by the CDC, underscoring its foundation in recognized public health research and planning methodologies (*Centers for Disease Control and Prevention (CDC)*, 2022; healthvermont, 2016).

### 5.5. Resilience in Community

"Community" commonly describes a diverse group of individuals residing in a shared geographic area. Bound by mutual interests, they engage in dynamic socio-economic interactions and collective action (Alshehri et al., 2015; Frankenberger et al., 2013; MacQueen et al., 2001;

Miles, 2015). However, this definition is not rigid; community boundaries can evolve or overlap due to advances in mobility and communication technologies. Moreover, individuals can concurrently belong to several communities, including those nested within larger ones (Mulligan et al., 2016; Sharifi, 2016).

Strategic Plan (Agency (FEMA), 2019), resilience means how well a person or community can endure, adjust, and bounce back from tough situations like natural disasters while keeping damages low. It involves being ready in advance, which includes reducing potential dangers, understanding the specific risks in their area, working together with relevant partners, and creating a culture that understands and responds to the different needs of the communities. This helps communities recover more effectively after a disaster. Community resilience, particularly in disaster preparedness and recovery, is intertwined with social capital, connectedness, equity, trust, community co-development, and feasibility. Social capital, derived from intra-community relationships and equitable practices, bolsters resilience by fostering inclusivity and engaging marginalized groups in planning and mitigation. Trust is a cornerstone; it encourages community participation and aids in making informed decisions for disaster risk mitigation. It is essential that this trust exists not only within the community but also between the community and external entities such as government agencies. Additionally, community co-development and ownership are fundamental in resilience-building processes, enhancing community commitment and accountability, essential for the long-term sustainability of resilience efforts.

Economic resilience, specifically, is integral to a community's overall resilience. It encompasses the capacity of the local economy to prevent, withstand, and recover from various economic shocks, including those that may arise from natural disasters, financial crises, or pandemics (Rose, 2004). Furthermore, there is a discernible nexus between economic resources and post-disaster well-being, particularly from the vantage point of social stratification. Socio-ecological studies incorporate resilience as "The ability of human communities to withstand external shocks or perturbations to their infrastructure, such as environmental variability or social, economic or political upheaval, and to recover from such perturbations" (Adger, 2000b). Economic resilience is underpinned not just by the magnitude but also by the diversity of economic resources. This broad understanding of economic resilience, encompassing growth, stability, diversity, and equity, is essential in ensuring that communities can recover from disasters and thrive in the face of ongoing changes and challenges.

Moreover, community resilience measures must be practical, affordable, and implementable at the grassroots level, considering limited community resources. Understanding these factors and their interplay is crucial for developing comprehensive strategies to bolster community resilience in the face of disasters or crises. Research plays a vital role in shedding light on these intricate dynamics and guiding the design of interventions tailored to the distinct needs and characteristics of each community (Engineering & Medicine, 2021).

As part of the ongoing efforts to build a comprehensive methodology for assessing community resilience, The National Institute of Standards and Technology (NIST) (Walpole et al., 2021) has developed a resilience indicator inventory, examining 56 resilience frameworks, as a valuable resource for practitioners and researchers in community resilience. NIST addresses the complexity of measuring community resilience, theoretical approaches, indicators, data needs, and scales of application observed in previous studies such as (Adger, 2006; Bakkensen et al., 2017; Cutter, 2016; M. Dillard, 2021; M. K. Dillard, 2017; Edgemon et al., 2018; Kwasinski et al., 2016, 2017; Lavelle et al., 2015; Serfilippi & Ramnath, 2018; Sharifi, 2016).

Figure 11 summarizes different focus areas in resilience definition. Agencies like The US Department of Homeland Security (DHS) and NIST have led the development of numerous methodologies, guidelines, and tools to strengthen community resilience. Similarly, FEMA has embraced the concept of community resilience indicators analysis in their reports (Gilbert et al., 2015; Security, 2015).

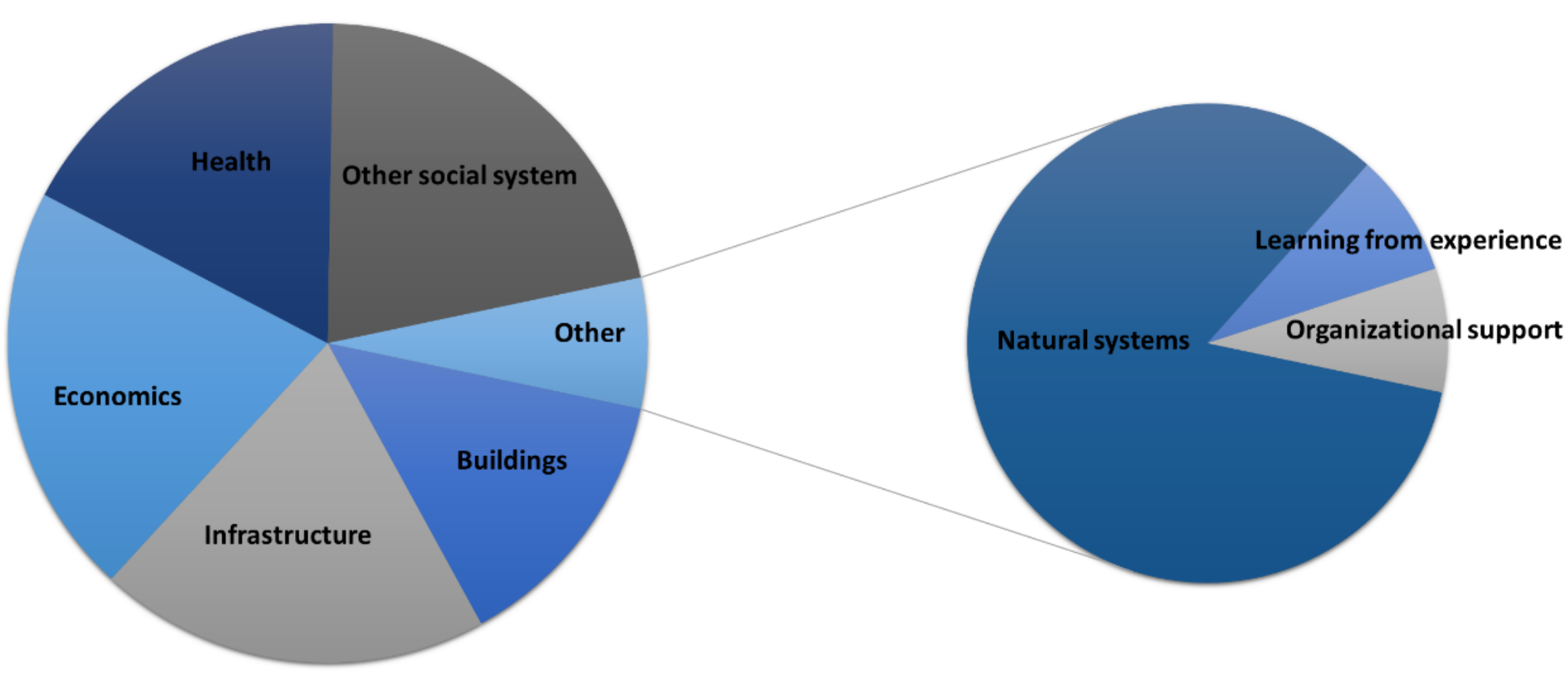

Figure 11. Focus areas in community resilience (Walpole et al., 2021)

The Federal Emergency Management Agency (FEMA) analyzes the community resilience indicators Analysis as process in their report Community Resilience (2022) and describes the Community Resilience Indicator Analysis (CRIA) process. The Joint Research Center (JRC) has developed the pioneering Geospatial Risk and Resilience Assessment Platform (GRRASP) in Europe. This platform utilizes geospatial technologies and computational tools to model resilience assessment for critical infrastructure. Concurrently, many initiatives, such as the 100 Resilient Cities (100RC) founded by The Rockefeller Foundation, aimed at enhancing global urban resilience to abrupt shocks like natural disasters and chronic stresses like unemployment and resource scarcity. Despite the program's conclusion in 2019, its legacy continues to inform urban resilience strategies worldwide, underscoring the need to adapt to evolving challenges like cyber threats and socio-political unrest (Spaans & Waterhout, 2017). Other resilience initiatives that improve community resilience include (Koliou et al., 2020; Pfefferbaum et al., 2013; Sempier et al., 2010). To address the prevention, protection, mitigation, response, and recovery of communities, individuals, families, businesses, local governments, and the federal government.

Nevertheless, a practical framework for assessing community resilience necessitates amalgamating physical infrastructure resilience metrics with social and economic systems (Kwasinski et al., 2016). As seen during Hurricane Katrina in 2005, the nexus allowing failures in one infrastructure, such as the outage of a particular electric power, could propagate to other infrastructures, often causing widespread disruption. The interplay between these interconnected infrastructures and community services greatly affects the resilience of the infrastructure systems, ultimately impacting the restoration of interdependent infrastructure networks (Karakoc et al., 2019).

### 5.5.1. Disaster Management

On Monday, July 3rd, 2023, the world experienced its hottest day on record. The U.S. National Centers for Environmental Prediction reported this unprecedented peak in global temperatures. This event surpassed the previous record set in August 2016, when the worldwide average temperature reached 17.01 degrees Celsius (62.62 degrees Fahrenheit). During that time, various regions around the world were grappling with intense heatwaves (Reuters, 2023). These increasing temperatures indicate the ongoing and intensifying phenomenon of climate change, which amplifies the frequency and severity of certain natural disasters, thereby necessitating a concurrent evolution in disaster management strategies. The emergence of Climate Change Adaptation (CCA) as a crucial strategy is a direct response to the multifaceted challenges posed by climate change, encompassing heightened global temperatures, a higher incidence of droughts, and an increase in storm intensity. These factors can lead to disaster or a radical transformation in the nature of disaster risk (*USGS*, 2022). CCA seeks to implement well-considered adjustments in both natural and human systems, responding proactively to climate change's confirmed or anticipated effects. The ultimate aim is to minimize negative impacts while maximizing any potential benefits that may arise (Change, 2007). Concurrently, Disaster Risk Reduction (DRR) plays a pivotal role in disaster management, with its core objective being to systematically analyze and address the causative factors of disasters. This encompasses initiatives aimed at reducing exposure to hazards, decreasing the vulnerability of people and property, promoting responsible management of land and environmental resources, and bolstering preparedness for potential adverse events (Reduction, 2009). Central to both CCA and DRR is the goal of mitigating natural disaster risks and enhancing community resilience (Turnbull et al., 2013). In an effort to combat the growing risks associated with climate change, the United Nations Office for Disaster Risk Reduction has introduced the Comprehensive Disaster and Climate Risk Management (CRM) program. This program is designed to weave climate information into the fabric of disaster risk management practices, aiding countries in their quest to anticipate better, prepare for, and mitigate disaster risks. This proactive stance is underpinned by alarming data, which indicates a 1.1°C increase in global temperature and a two-fold increase in extreme weather events over the last 20 years (*Comprehensive Disaster and Climate Risk Management*, 2022).

The implications of climate change-driven hazards are profound, with the potential to escalate disruptions and simultaneously erode the resilience of households and communities, a reality underscored by (*PreventionWeb*, 2022). Building resilience is crucial, especially in communities that are disproportionately affected by the risks associated with climate change, globalization, and urbanization. The Intergovernmental Panel on Climate Change (IPCC) defines resilience as the capacity of a system and its components to anticipate, absorb, accommodate, or recover from hazardous events efficiently while maintaining or improving essential functions (Field et al., 2012). Enhancing community resilience involves not only mitigating the adverse consequences of adversities but also empowering communities to recover swiftly and emerge stronger. The strategies aimed at achieving this encompass vulnerability reduction and resilience enhancement, which are central tenets shared by both CCA and DRR (Mitchell & van Aalst, 2008). Addressing these issues necessitates a holistic disaster management approach that spans preparedness, response, and rehabilitation. By effectively organizing, coordinating, and utilizing resources and responsibilities, disaster management strives to minimize the detrimental impacts of disasters and manage the ensuing consequences, facilitating a swift return to normalcy and reinforcing community resilience.

## 6. Conclusion

The concept of resilience, while universally acknowledged as vital in both research and practical applications, presents a complex challenge in terms of its definition and measurement across various disciplines. This complexity arises from the inherently multidimensional nature of resilience, which is interpreted and applied differently in fields such as social science, engineering, psychology, and community studies. Despite these varied perspectives, at its core, resilience encapsulates the capacity of a system—be it an individual, a community, a societal structure, or a physical infrastructure—to withstand, adapt to, and recover from disturbances. In Social Science, resilience is often viewed through the lens of societal and community capability to endure and adapt to social, economic, and political challenges. This perspective emphasizes the adaptive capacity of social systems and the role of human agency in navigating complex social dynamics. In Engineering, the focus shifts to the resilience of physical systems and infrastructures, highlighting their ability to resist, absorb, and recover from external shocks and stressors. This interpretation prioritizes structural integrity and the design of systems that can maintain functionality in the face of disruptions. In Psychology, resilience is primarily concerned with individual mental health and well-being. It revolves around the capacity of individuals to cope with stress and adversity, maintaining psychological equilibrium and the ability to bounce back from traumatic or challenging experiences. In Community Studies, resilience is often examined at the community level, encompassing the collective response and adaptation to various challenges. This includes the role of social networks, community resources, collective action, and shared resilience strategies. Despite the varying angles from which these fields approach resilience, there is a harmonious thread: the emphasis on the capacity to respond and adapt to challenges. However, conflicts arise in the specific focus and application of resilience concepts. For instance, while engineering may prioritize physical robustness, psychology might focus more on emotional adaptability and coping mechanisms. Integrating these diverse definitions contributes to a more comprehensive understanding of resilience, providing a multifaceted view that recognizes both individual and systemic aspects. This comprehensive approach is particularly important in the context of global challenges such as climate change, public health crises, and socio-economic instability, where the interconnectedness of systems becomes starkly apparent. However, gaps remain in the current body of literature, particularly in integrating and applying these varied perspectives. Future research should aim to bridge these gaps by developing a multidimensional framework of resilience that incorporates insights from different disciplines. This framework could lead to the development of standardized resilience definitions and metrics that are adaptable to the nuances of different fields while maintaining a core set of principles. Moreover, the increasing complexity of global systems, as exemplified by events like Hurricane Katrina, highlights the need for a more integrated approach to understanding and enhancing resilience. This includes examining the cascading effects of system failures and the interdependencies between different types of systems. Future research could significantly benefit from utilizing advanced data analysis techniques for risk assessment, understanding the role of social networks and leadership in resilience, and developing effective educational and training programs to enhance individual and collective resilience. By synthesizing these diverse perspectives, future research can contribute to developing more effective strategies for managing and mitigating risks, thereby enhancing the resilience of systems on a global scale. In conclusion, the pursuit of a unified understanding of resilience across various disciplines is not only a scholarly endeavor but a practical necessity. Our review aims to contribute significantly to this field, offering insights and directions that could inform practitioners and researchers alike in their ongoing efforts to comprehend and enhance resilience in an increasingly complex world.